\documentclass[12pt]{iopart}

\usepackage{graphicx}
\usepackage{color}

\begin{document}

\title[\textsl{In situ} compensation method for ... ]{\textsl{In situ} compensation method for high-precision and high-sensitivity integral magnetometry}

\author{Katarzyna Gas and Maciej Sawicki}

\address{Institute of Physics, Polish Academy of Sciences, Aleja Lotnikow 32/46, PL-02668 Warsaw, Poland}

\ead{kgas@ifpan.edu.pl, mikes@ifpan.edu.pl}

\vspace{10pt}
\begin{indented}
\item[02 ]May 2019
\end{indented}

\begin{abstract}
The ongoing process of the miniaturization of spintronics and magnetic-films-based devices, as well as a growing necessity for basic material research place stringent requirements on sensitive and accurate magnetometric measurements of minute magnetic constituents deposited on large magnetically responsive carriers. However, the ever so popular commercial integral magnetometers based on superconducting quantum interference device sensor are not object-selective probes. Therefore, the sought signal is usually buried in the magnetic response of the carrier, contaminated by signals from the sample support, system instabilities and additionally degraded by inadequate data reduction.
In this report a comprehensive method based on \textsl{in situ} magnetic compensation by sample abutting long strips made of matching material is shown to provide means for mitigating all these weak elements of integral magnetometry.
Practical solutions and proper expressions to calculate the absolute values of the investigated moments are given. Their universal form allows to employ the suggested design in investigations of a broad range of specimens of different sizes, shapes and compositions. The method does not require any extensive numerical modeling, relying only on the data provided by the magnetometer. The solution can be straightforwardly implemented in every field where magnetic investigations are of a prime importance, including the emerging new fields of topological insulators, 3D--Dirac semimetals and 2D--materials.
\end{abstract}

%
%
%
%
%

\section{Introduction}

Since the advent of commercially available, affordable, automated (computer controlled), basic fault--tolerant and user-friendly superconducting quantum interference device (SQUID) integral magnetometers in the early 90s, sensitive SQUID-based magnetometry has now established itself as an indispensable everyday characterization and experimental tool in modern science. 
It has become routinely performed on a wide range of different types of samples such as ultrathin films \cite{Ney:2001_EPL,Sawicki:2010_NP,Chibal:2016_SR,Gladczuk:2017_JPDAP,Hayassi:2018_APE,Sawicki:2018_PRB}, nanoparticles \cite{Sandaresan:2006_PRB,Sadowski:2011_PRB,Sueli:2016_JPCC,Rath:2011_JMMM,Augustyns:2017_PRB}, nanowires \cite{Siusys:2014_NL,Sadowski:2017_Nanoscale},
nanocomposites \cite{Peddis:2008_JPCC}, quantum dots \cite{Sun:2017_QM}, graphene \cite{LiuYuan:2013_SR,LiuYuan:2016_NC},
dilute magnetic \cite{Pereira:2011_JPCM,Sawicki:2013_PRB,Pereira:2013_JPCM,Henne:2016_PRB,NeyV:2016_PRB},
and ferromagnetic semiconductor epilayers \cite{Sawicki:2006_JMMM,Gas:2018_JALLCOM}.
It currently enters the realms of organic spintronics \cite{Bujak:2013_CSR}, biology \cite{Kopani:2015_BioMetals}, and topological matter \cite{Zhao:2014_NM,Dutta:2017_SR}.
Despite this broad range of subjects, the investigated materials share one crucial common aspect: the objects of interest come on bulky substrates, or at least have to be fixed to a kind of rigid carrier, permitting mounting them onto adequate sample holders suitable for withstanding measurement cycles 
in wide ranges of temperature $T$ and magnetic field $H$.
However, the most popular multipurpose commercial SQUID magnetometers are not object- or element-selective probes, so the sought signal is very often (deeply) buried in the magnetic response of the ("nonmagnetic") carrier.
The latter becomes dominant in the mid- to strong-magnetic-field range, when the signal from the investigated sample usually saturates, but that of the carrier continues to grow.
To make things worse the signals exerted by such complex objects are detected and processed with some limited accuracy by the magnetometer hardware and software and finally can be, frequently, mishandled by a less experienced end user.
Here, a lack of a proper laboratory procedure which may lead to magnetic contamination \cite{Abraham:2005_APL,Garcia:2009_JAP,Pereira:2011_JPDAP,Sawicki:2011_SST,Ney:2011_SST} and other experimental artifacts \cite{Ney:2008_JMMM,Pereira:2017_JPDAP,Buchner:2018_APL} has to be underlined.
The latter can be sizably enlarged when the operator lacks the general recognition of how the measurement process and data processing and reduction is carried out \cite{Sawicki:2011_SST}.

It has to be noted here that mitigating the commonly met problems, which is an indispensable step in responsible (SQUID) magnetometry, does not necessarily lead to the reduction of the \textsl{real} magnitude of the error bar towards the typically declared sensitivity by the manufacturers (usually around $10^{-8}$~emu).
Such a precision indeed can be required in studies of 
very thin layers of magnetically diluted compounds or antiferromagnets, particularly when their magnetic anisotropy is  targeted.
The same challenge is faced when the critical exponents specific to magnetic phase transitions in very thin films are to be determined \cite{Stefanowicz:2013_PRB,WangMu:2016_PRB}.
It has been established, following the best guidance and rules put forward to date in the literature, \cite{Stamenov:2006_RSI,Ney:2011_SST,Sawicki:2011_SST,Pereira:2017_JPDAP}
that the most credible assessments of the sought moment $m_{\mathrm{X}}$ can be obtained upon a subtraction
of the results of  two separate measurements: that of the sample $m_{\mathrm{S}}$ and of its support $m_{\mathrm{R}}$, usually the bare substrate or the carrier.
However, because both responses are mostly composed of the signal from the support, both the minuend and the subtrahend are very large numbers of comparable magnitude.
The key point here is that the credibility of a difference between two such numbers directly depends on the absolute precision and accuracy with which \textsl{both} these numbers were established.
Therefore, even minute magnetometer instabilities occasionally affecting the results of the measurement process make the required subtraction substantially erroneous.
This in turn degrades the scientific soundness of the established results - frequently without issuing any clear warning sign to the researcher.

In this report we put forward a carefully elaborated and experimentally validated experimental approach which increases the reliability of the above mentioned subtraction method by sizably reducing the vulnerability of the final results to the randomly appearing disturbances in the magnetometer's output.
The provided below exact formulae allows vastly advance both the precision and reliability of the experimentally established signals of very weak magnetic sources which come on bulky substrates.
Importantly, this approach is fully compatible with typically employed experimental methodology used on the everyday basis in magnetometry labs.

Our report is organized as follows.
First, we point out and exemplify the magnitude of the main system instabilities and spurious components contributing to the magnetometers' outputs which cannot be eliminated by the virtue of the subtraction method.
Then we introduce the basics of the \textsl{in situ} compensation approach and detail the experimental method up to the point where the final expression for the desired moment is given.
The report is ended by presenting examples of the sizable improvement in the consistency of results obtained by employing the compensational approach and by the final conclusions.

\section{Equipment-related limitations to precise magnetometry}

The study reported here is based on years of experience gained working with Quantum Design Magnetic Property Measurement System  (MPMS) SQUID magnetometers.
In the authors' view there are two most obvious \textsl{equipment-related} reasons which can affect not only the magnitude of the established magnetic moment $m$, but can also corrupt its dependence on $H$ and $T$, $m(H)$ and $m(T)$, respectively.

The first problem arises from a non-perfect performance of the sample transport mechanism during scanning, that is, tripping the sample up and down along the SQUID pick-up coil axis $z$.
This is the moment when the experimental response function $V(z)$, i.e. the dependence of the SQUID response \textsl{vs}. sample position, is established.
It is important to be reminded here, that the magnitude of $m$ is established by a least-mean-square fitting of a reference $\Upsilon(z)$ function into $V(z)$.
$\Upsilon(z)$ describes the response of the magnetometer's sensing coils to the position of an ideal point dipole of a constant magnetic moment along the $z$ axis.
However, the MPMS system does not measure the actual position of the sample during scanning.
The system only assigns a certain value of $z$ according to the time interval passed from the beginning of the scan.
Therefore, any mechanical wear and tear in any of the mechanical components of the sample transport mechanism mars the shape of $V(z)$, and imprints on the magnitude of the established $m$.
Obviously mechanical effects accumulate over time, and frequently evade the attention of the user(s).

The main distortions of $m(H)$ originate from only approximately correct magnitudes of $H$ which are reported in the magnetometer's output files.
This inaccuracy stems from the fact that the commercial SQUID magnetometers do not contain any built-in field sensors.
Therefore, instead of the real magnitude of the magnetic field $H_{\mathrm{real}}$ - the field experienced by the sample during the measurement, only the value requested by the user $H_{\mathrm{set}}$ is reported.
The point is that $H_{\mathrm{set}}$ only occasionally precisely corresponds to $H_{\mathrm{real}}$.
There are many sources of $H_{\mathrm{set}} \leftrightarrow H_{\mathrm{real}}$ discrepancy \cite{Sawicki:2011_SST}.
The main one for these considerations seems to be related to a limited accuracy of the digital-to-analog converter which controls the magnitude of the output current $J_H$ of the magnet power supply.
In addition, the magnitude of $J_H$ may vary in response to an unstable environment, for example to variations of the ambient temperature in the lab.

On the other hand, it has to be underlined that  nowadays the available commercial magnetometers are very well engineered and are generally very well suited for the majority of tasks.
It is only the necessity to assure a very high precision and reproducibility of $m$ that points to the above mentioned equipment insufficiencies.
As shown below their presence precludes a fulfilment of some of the very demanding research tasks without employing more advanced measures to reduce the influence of the above mentioned main weaknesses in the design or performance.

In order to exemplify the scale of the challenge addressed here and to show to what extent such issues can be mitigated, the measurements of very thin layers of the dilute magnetic semiconductor (Ga,Mn)N are considered.
This magnetic form of GaN in which a small percentage of Ga atoms is randomly substituted by Mn, belongs to increasingly important group of Rashba materials \cite{Stefanowicz:2014_PRB,Adhikari:2016_PRB}.
(Ga,Mn)N compounds are typically deposited by molecular beam epitaxy  method \cite{Kunert:2012_APL,Gas:2018_JALLCOM} on GaN-buffered (2 -- 3~$\mu$m) sapphire substrates (300 -- 500~$\mu$m thick), whose strong diamagnetic response overwhelmingly dominates the signal of the dilute magnetic layer, particularly at large $H$ and for layers for which the thickness $d$ is in the single nanometers range.
The problem of the correct extraction of the layers' signal remains sound even at very low temperatures, where high quality single phase (Ga,Mn)N layers exhibit low temperature ferromagnetic properties \cite{Sawicki:2004_ICPS,Sawicki:2012_PRB,Kunert:2012_APL,Gas:2018_JALLCOM}.
The precision is of a great value here since, owing to the $L = 2$ and $S = 2$ configuration of the Mn$^{3+}$ species a field of the order of a few tens of kOe is required to reach the sufficient level of saturation, which would permit an establishment of Mn concentration $x$ with a satisfactory accuracy.
It has to be noted that magnetic characterization by SQUID magnetometry is practically the last resort enabling a reasonable estimation of $x$ for a mere nanometre thin and/or magnetically diluted layers ($x < 0.1$).
However, caution is required, the precision of the magnetometric determination of $x$ can be compromised by a lack of precise knowledge about the Mn oxidation state. 
A synchrotron based X-ray absorption near-edge spectroscopy method appears to be the most suitable for this purpose \cite{Stefanowicz:2010_PRB,Bonanni:2011_PRB,Gas:2018_JALLCOM}.

To put the things into perspective we note that an expected magnitude of the saturation moment $m_{\mathrm{sat}}$ of a typical $\sim 5 \times 5$~mm$^2$ piece of $d=5$~nm (Ga,Mn)N layer containing about 5\% of Mn is about $5\times 10^{-6}$~emu or 5~$\mu$emu.
Therefore, to stay below a decent 10\% relative error bar for the Mn concentration $x$ (it is the accuracy which is typically required as feedback information to the sample growers) the magnitude of $m_{\mathrm{sat}}$ in such layers has to be established with the absolute accuracy of better than 0.5~$\mu$emu, well above the manufacturer's declared floor level of 0.01~$\mu$emu.
However, this simple evaluation does not mean one is safe here.
On the contrary, the situation changes completely when two additional factors are taken into account.
First is the presence of a bulky substrate, second that a large range of strong fields, $40 < H < 70$~kOe, is needed to assess $x$ reasonably well \cite{Bonanni:2011_PRB,Gas:2018_JALLCOM}.
At this $H$-range  a relevant 0.33~mm thick sapphire substrate exerts a flux corresponding to about -1000~$\mu$emu, so $m_{\mathrm{S}}(H)$ of both the whole sample (the layer plus the substrate) and of the bare substrate [the reference $m_R(H)$] must be established with the same absolute accuracy better than 0.5~$\mu$emu.
This actually means that both these dependencies must be known with an absolute precision exceeding 0.05\%, or 3 1/2 digits.
Evidence is given below that a typical MPMS system is not capable of maintaining such a stability of its output in a time frame sufficiently long to permit completion of the whole set of the required measurements.

Far more severe constraints are imposed by studies of the magnetic anisotropy.
It is of a single ion origin in (Ga,Mn)N and it does not "saturate" at any field attainable in SQUID magnetometers \cite{Stefanowicz:2010_PRB,Sztenkiel:2016_NC}.
It actually means that at low temperatures magnetic moments established along both hard and easy axes (perpendicular and in-plane orientations, respectively) do not fall on top of each other
even at $H = 70$~kOe - the maximum value of $H$ attainable in commercial SQUID magnetometers.
Yet, the magnitude and the field dependence of the magnetic anisotropy provides an indispensable reference for the theoretical description of the magnetism in this \cite{Stefanowicz:2010_PRB,Sawicki:2012_PRB,Sztenkiel:2016_NC} or similar material systems \cite{Sawicki:2013_PRB}.
In this case the magnitudes of the high field moments measured in both these orientations should be known with an absolute accuracy better than 0.1~$\mu$emu, or full 4 digits.

\begin{figure}[t]
	\includegraphics[width=0.9\columnwidth]{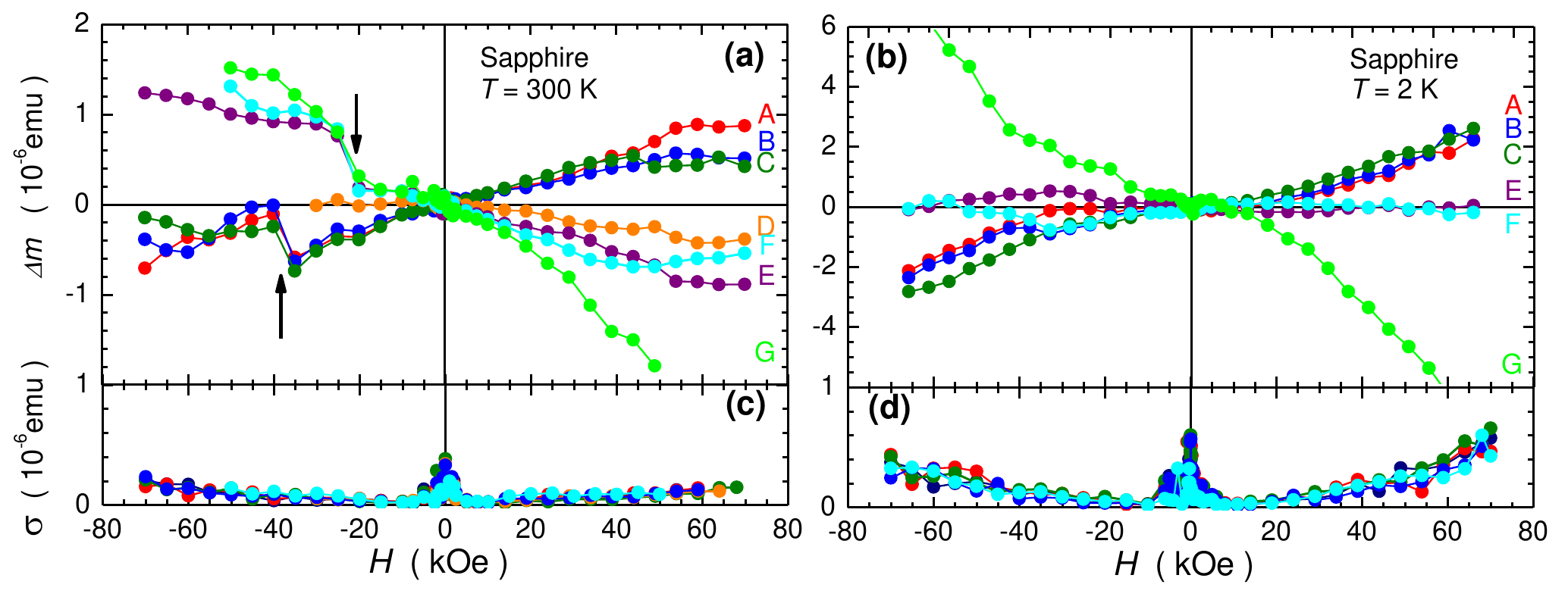}
	\caption{\label{fig:1-SzBrA}%
Magnitudes of typical instabilities of MPMS output exemplified by a comparison of the $m(H)$ dependencies obtained for the same reference $5 \times 5 \times 0.33$~mm$^3$  sapphire sample at 300 and 2~K, panels (a) and (b), respectively. The plotted differences $\Delta m(H) = m(H) - m_{avg}(H)$ are calculated for each $m(H)$ taking an average $m_{avg}(H)$ of all the $m(H)$ presented on the panel as the reference. Curves A -- E were obtained  in a short time frame (one month), curve F a year later, and the curve G after two years from the first measurements. These measurements have been made using conventional sample holders, identical to that shown in Fig.~\ref{fig:CSH}~(a).
The arrows indicate events of sudden shifts of $m$ ("jumps").
For the clarity of presentation the corresponding standard deviations $\sigma$ for these points are given in separate panels: (c) for 300 K and (d) for 2 K.
The sizable magnitude of $\sigma$ at weak field is connected with a strong magnetic flux creep and escape in the superconducting magnet \cite{QD1070-207-MST}.
The same $m(H)$ are presented in a different manner in Supplementary Figure S1.}%
\end{figure}
The scale of the typical output instabilities of a well maintained MPMS systems is exemplified in Figure~\ref{fig:1-SzBrA}, where results of $m(H)$ measurements of \textsl{the same} reference piece of a sapphire substrate acquired within a period of nearly two years  are compared.
Plotted in panels (a) and (b) are the differences $\Delta m(H) = m(H) - m_{avg}(H)$ calculated for each $m(H)$ taking an average $m_{avg}(H)$ of all the $m(H)$ from each panel as the reference.
The aim of all these measurements was to validate, and, as it turned out -- to update, the substrate reference $m_{\mathrm{R}}(H)$, which must be subtracted from the samples' $m_{\mathrm{S}}(H)$ in order to establish $m_{\mathrm{X}}(H)$. 
As the data clearly illustrate for both $T = 2$ and 300~K, the observed instabilities ($\Delta m \simeq 1$~$\mu$emu) at high field only barely allow for basic magnetic characterization of very thin dilute films, and completely preclude investigations of their magnetic anisotropy.

Another troublesome element of the results presented in Figure~\ref{fig:1-SzBrA} is the presence of "jumps" or "kinks".
The most pronounced ones are indicated by the arrows in panel (a).
It has been checked that the scans acquired before and after such an event are technically of the same high quality, which eliminates a systematic distortion of $V(z)$ as the culprit.
This is further confirmed by a smooth behaviour of the standard deviation $\sigma$  at these regions of $H$ where the irregularities occur.
We therefore assign their presence to minute inaccuracies in the field setting channel of the magnetometer.
Actually, to reproduce the reported  magnitude of the "jumps" of about 0.5~$\mu$emu at $H \simeq$~-40~kOe, a suddenly appearing difference between $H_{\mathrm{real}}$ and $H_{\mathrm{set}}$ of about 40~Oe is sufficient.
 This is only about 0.1\% of the intended magnitude of $H$.
It has been observed that these two features affecting $m(H)$ come and go with different intensity and that they do not correlate either with a change of the noise level, or with other disturbances in the lab or its vicinity, or with a particular time of the day.

\begin{figure}[t]
	\includegraphics[width=0.9\columnwidth]{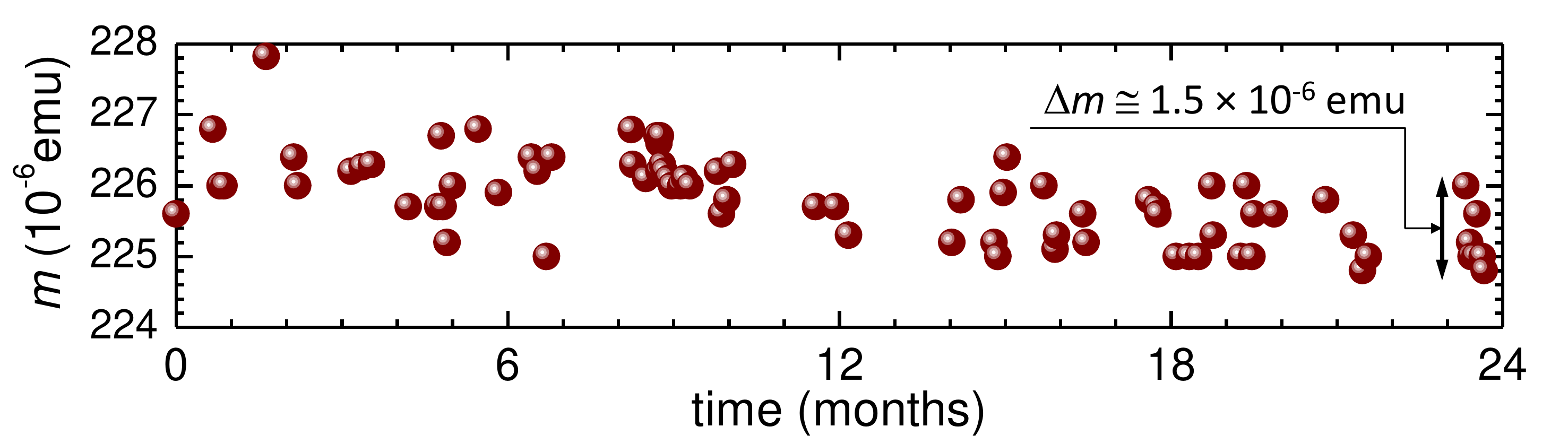}
	\caption{\label{fig:Czas} %
MPMS SQUID magnetometer instability. All the measurements have been taken at nominally identical conditions ($T = 300$~K and $H = 20$~kOe, the latter ramped always up from zero) during the centering of one compensational sample holder (CSH) considered in this report [depicted in Figure~\ref{fig:CSH}~(c)].
In this case a 5~mm wide gap between sapphire compensating strips serves as the source of the reference signal.
The time scale on the $x$ axis is counted from the date this CSH has been tested for the first time.
The double arrow indicates the magnitude of the short time spread of the points (obtained, say, within 1-2 weeks), $\Delta m \simeq 1.5$~$\mu$emu, which matches well the observed spread of $\Delta m$(20~kOe) seen in Figure~\ref{fig:1-SzBrA}~(a).
}%
\end{figure}

Another estimation of the MPMS's output instability comes from a set of systematic measurements of a reference signal source at nominally identical settings.
The interrogated object in Figure~\ref{fig:Czas} corresponds very well to the sapphire substrates considered in Figure~\ref{fig:1-SzBrA}. 
The main message from this two years long test is that although the long time average of the tested MPMS magnetometer stays at reasonable consistent level, at a short time frame, say 7-10 days, the output fluctuates with disappointingly high magnitude of $\Delta m \simeq 1.5$~$\mu$emu at 20~kOe and
possibly at 5~$\mu$emu after extrapolation to 70~kOe.
According to the estimates presented before, both these values are \textsl{prohibitive} for the reliable establishment of minute magnetic signals
basing only on the subtraction approach.

Two important conclusions can be drawn from these tests.
Firstly, that in the subtraction method one cannot rely on a "universal background curve".
Particularly, when the investigations are carried over an extended time frame.
Secondly, that this is the magnitude of the output instability that is responsible for the sizable reduction of the attainable sensitivity of the SQUID magnetometry, even if it is performed in accordance with the best known experimental protocols.
These two points immediately eliminate from consideration the automated background subtraction method provided by MultiVu software, since it also establishes the magnitude of $m_{\mathrm{X}}$ upon a difference of, essentially, same $m_S$ and $m_R$ measurements.

It has therefore  become obvious that in order to push down the limits of reliable magnetometry in the existing magnetometers one has to significantly reduce the contribution from the carrier of the researched object.
However, for a day-to-day characterization and related basic research the mechanical or chemical thinning of sapphire down to the single $\mu$m range has not been a feasible option.
Therefore, a method of an active elimination of the substrate signal remained the only viable and affordable solution.
In the following parts of this report the basic concept of an active, \textsl{in situ} type of, substrate compensation method is presented, detailed, quantified and put to the test.

\section{Principles and challenges of the \textsl{in situ} compensation}

The underlying concept of the \textsl{in situ} compensation of the spurious signal of the substrate arises from the fact that in these magnetometers in which the detection of the magnetic moment relies on (an axial) movement of the sample for a scan length of a few centimeters with respect to the self-balanced sensing coils, any infinitely long shape made of magnetically \textsl{homogeneous} material (even an iron bar) will not produce any signal.
This stems from the fact that for each individual fragment of the infinite rod there exist a partner which produces a response of the same magnitude but of the opposite sign.
In practice the phrase "infinitely long" can be replaced by "sufficiently long", with the required length becoming shorter the less magnetic is the considered material.

\begin{figure}[t]
	\includegraphics[width=0.7\columnwidth]{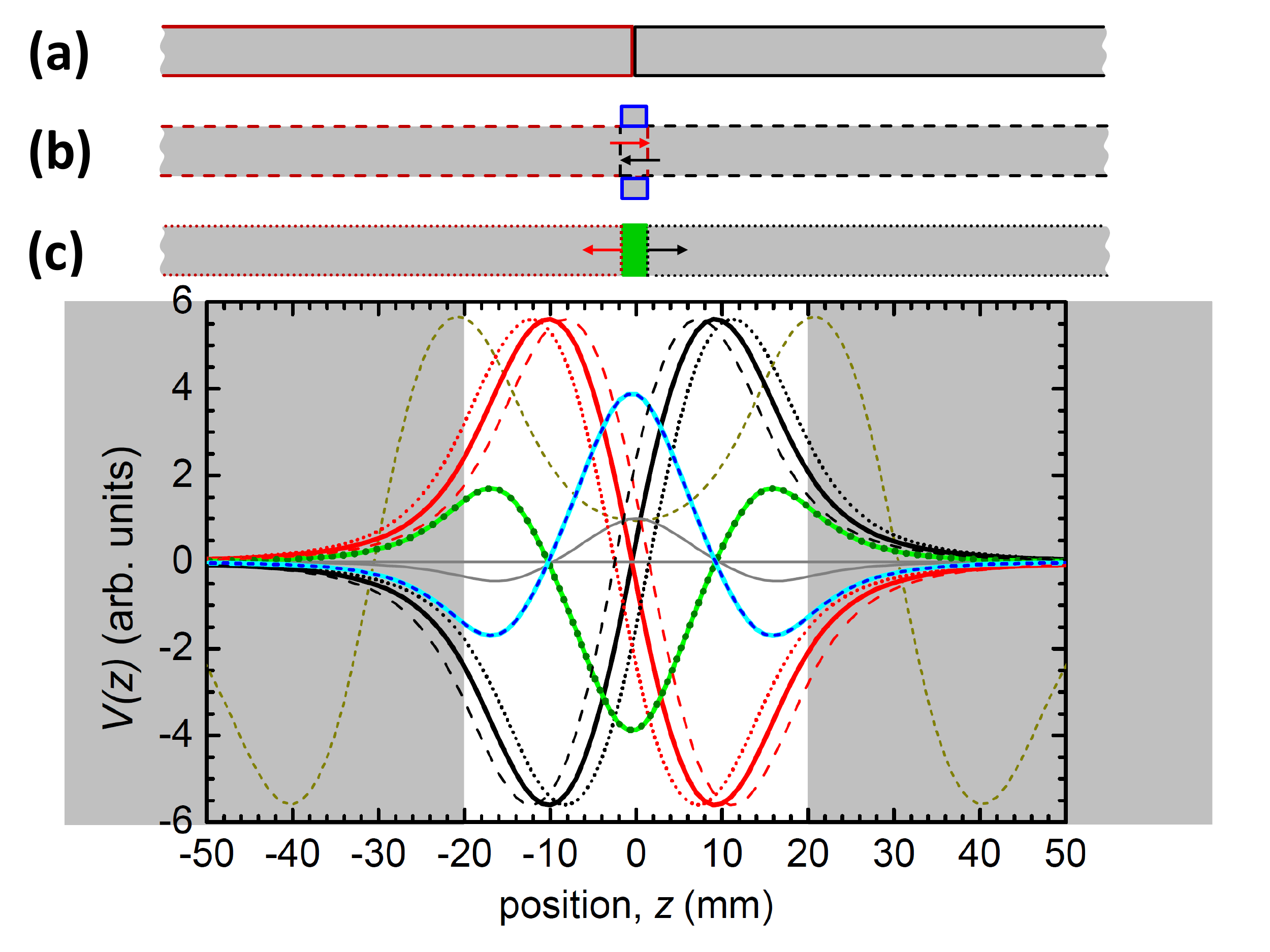}
	\caption{\label{fig:ES(ends)} %
	(Main panel) Simulations of the changes of the signal $V$($z$) picked up by the $2^{nd}$ order gradiometer for the three cases (\textbf{a} -- \textbf{c})
depicted in upper part of the figure. The solid red and black lines in the graph represent individual responses of semi-infinite rods cut from a uniformly magnetized material at position $z = 0$, case (a). If moved together, the resulting $V(z) = 0$ everywhere. By shifting these halves from $z = 0$, by 2 mm against each other - case (b), their responces shift by the same amount, dashed lines in the graph of matching colors, so their sum acquires sizable amplitudes near the center (the blue dashed line), or near the bulge created by the overlapping rods. In fact the case (b) corresponds to the case (a) of an infinite rod with a muff located at its center (outlined in blue). This reflects the standard situation when samples are mounted on a very long sample holder. Indeed, the dashed blue sum corresponds precisely to the computed $V(z)$ of a 4~mm fragment of this rod, i.e. a sample, when it is tripped alone through the pick-up coils (the light blue background for the dashed blue sum). By pulling these halves away from each other, say, again by 2~mm each way from their joining point, case (c), a 4~mm wide gap is formed (green). Its magnetic response (dotted dark green line) can be obtained again as the sum of the individual $V(z)$ of both halves after shifting them to their new starting positions (dashed lines of matching colors).
Importantly, this gap-related $V(z)$ has got exactly the same form, but is of an opposite sign to the sample-like $V(z)$ calculated for the case (b).
This effective mutual cancelation of cases (b) and (c) constitutes the basis of the active \textsl{in situ} compensation of chunks of supporting material (substrates)
accompanying the minute objects of investigations.
The dark yellow short-dashed line exemplifies a response of only a 60~mm long rod. It corresponds to the approach exercised in ref.~\cite{Fukumura:2001_APL}.
The thin grey solid line indicates the ideal point-dipole $V(z)$ of the material from which the rods are made and its maximum value $V(0)$ serves as the reference level for all the other curves plotted in this graph.
In the considerations presented here the effects of a radial extent of the objects on $V(z)$ has been neglected.
We touch upon this issue in section S4 of the Supplementary Information. 
The white background box marks the extent of the typical scanning window, when a high precision of the measurements is required.}
\end{figure}

Actually, the influence of a length of a specimen on the magnetometer's response can be viewed as an "ends issue".
It is illustrated in Fig.~\ref{fig:ES(ends)}.
We start our considerations from an "infinitely long" homogenously magnetized rod which is divided in its center into two semi-infinite parts [Fig.~\ref{fig:ES(ends)}~(a)].
In the main panel of Fig.~\ref{fig:ES(ends)} we plot the calculated SQUID responses $V(z)$ of each individual half-rod, taking $z = 0$ as the center of the sensing coils (red and black solid lines for the left and right halves, respectively).
Throughout the whole study we assume the sensing coils to be arranged in the second order gradiometer fashion, as in the MPMS units, and for the numerical simulation we take the gradiometer's coils dimensions according to MPMS specification~\cite{QD1014-213-MST}.
It is clearly seen, that the strongest signal of the individual half-rods are "generated" at the close vicinity to their origin, that is, at the place where the symmetry of the semi-infinite rod is the lowest.
Such a resonant-like shape of $V(z)$ is the result of the particular gradiometer configuration and the two extrema existing on each $V(z)$ occur at the positions $z = \pm R$, where $R=9.7$~mm in MPMS magnetometer is the radius of the gradiometer coils.
Actually, this is the most general rule for this kind of magnetometry - the strongest signals come from these places where the translational symmetry is reduced.
Most importantly, the sum of these two $V(z)$ nulls out everywhere, as expected for an infinite homogenous object.

\label{sec:basics}

By exactly the same token the system picks the signal from the sample - a typical short specimen can be viewed as two symmetry breaking ends located very close to each other (at the sample length distance), whose signals mutually cancel out away from their ends (i.e. the sample).
This situation is simulated in the panel (b).
Here, by pushing the same halves against each other, say by 2~mm from their origin, a 4~mm long bulge is created in the overlapping area,
outlined in blue.
And it is only the 4~mm long muff located symmetrically with respect to the center that breaks the translational symmetry of the whole structure, so its flux  nulls out in the pick-up coils except only from the muff.
The corresponding $V(z)$ for each of the halves are marked in dotted lines of matching colors and their sum, the dotted blue line, indeed nulls everywhere except of the (relatively) narrow region in the vicinity of the muff.
Importantly, it is exactly the same $V(z)$ which will be induced by a 4~mm long fragment of this rod, i.e.~a sample, when tripped alone through the pick-up coils.

The dotted grey line in Fig.~\ref{fig:ES(ends)} exemplifies the ideal point-dipole $V(z)$ of the material from which the rod is made of and its maximum value $V$(0) serves as the reference level for all the other curves plotted in this graph.
And the comparison of the amplitudes of this $V(z)$ and that of the 4 mm long muff explains why in the conventional
magnetometry it pays to increase the linear dimensions of the investigated (homogenous) samples - it reduces the effect of flux cancellation.
Certainly, the net gain in the effective flux due to extending the length of the specimen has its limit (it has to null again for the infinitely long sample) and reaches its maximum when the length of the sample approaches $R$.\cite{Hayden:2017_RSI}
More general considerations concerning the influence of samples' linear dimensions and their alignment can be found elsewhere \cite{Zieba:1993_RSI,Miller:1996_RSI,Stamenov:2006_RSI,Sawicki:2011_SST}.

Both halves of the rod can also be pull apart from their joining point at the center, say, by the same 2~mm.
It is illustrated in panel (c).
In this case a 4~mm gap is formed (marked in green), again sizably breaking the symmetry of the rod near the center.
The corresponding $V(z)$ for the separated semi-rods are marked by dashed lines in the graph.
As in the previous case, the sum of the two dependencies (green solid line) nulls everywhere except of the
(relatively) narrow region close to their ends.
Most importantly, this sum corresponds precisely to a $V(z)$ of the symmetry breaking muff, but is of the opposite sign.
So, both cancel each other completely.
This fact constitutes the essence of the \textsl{in situ} compensation method: a long sample holder made with an adequate material has to posses a gap in which the investigated sample can be nested - ideally filling the gap completely.
In this arrangement, the flux of such artificially elongated substrate will get nulled, 
leaving only the flux of the layers of interest for the detection.
The latter can now be established with a far greater precision since the symmetry breaking by the layer is sizably stronger than the symmetry breaking by the elongated substrate.

A very elegant attempt to realize the \textsl{in situ} compensation  was proposed by T. Fukumura \textsl{et al.} \cite{Fukumura:2001_APL}, in their study of MBE grown ZnO epilayers with Mn.
In order to eliminate the magnetic signal from the sapphire substrate, a mere few mm of the central part of about 60~mm long and 5~mm wide sapphire substrate strip was overgrown by about 2~$\mu$m thick (Zn,Mn)O film.
As presented in Fig.~\ref{fig:ES(ends)}, in such a geometry only 70 -- 80\% of the substrate contribution could had been compensated.
Nevertheless, the arrangement proved sufficient to observe the positive signal due to Mn ions in wide temperature range and in $H$ up to 50~kOe.
Although very instructive and somehow effective, this solution is of a little practical value in typical laboratory practice, since most materials are deposited uniformly on large surfaces (substrate availability depending) and small specimens are cleft for further characterization or investigations.

A more practical approach was implemented by M. Wang \textsl{et al.} \cite{WangMu:2016_PRB} for accurate studies of the critical region of the Curie transition in (Ga,Mn)As.
This compound is grown on GaAs and the investigated $5 \times 5$~mm$^2$ samples were cleft from standard 2" substrates.
For the SQUID magnetometry the samples were abutted on either side with strips of GaAs of the same width as the sample.
All three pieces were glued to a length of 5~mm wide Si supporting stick.
According to those authors both the Si support and the GaAs strips extended much further than the length of the detection coils, making the magnetometer effectively insensitive to the magnetic flux due to the layer's substrate and the Si support.
The concept indeed proved very efficient in eliminating the spurious signal of the GaAs substrate.
The authors assumed a perfect compensation in their particular case and did not consider a more general situation of a not perfect match of the sample to the abutting strips.
This is, however, very likely because of the spread of substrates' thicknesses -- typically up to  30~$\mu$m, or 2 -- 5\% relative, even for wafers originating from the same batch.
Such discrepancies could not have produce any noticeable signal in the very low field ($H \leq 300$~Oe) magnetometry exercised in \cite{WangMu:2016_PRB}.
This is simply because in such a weak fields region the signals related to a small imbalance fell below the magnetometer's ultimate sensitivity.
But in the high--$H$ region the situation changes completely.
In section \ref{sec:wzor} we detail the method how to extract the moment of interest in the case of an arbitrary \textsl{in situ} compensation, which works at any $T$ and $H$ within the range available in the magnetometer.

Another approach of background subtraction is based on a differential approach \cite{Cabassi:2010_MST}.
It strictly relies on a perfectly symmetrical mounting of the sample and the relevant reference.
However, the final $m_{\mathrm{X}}$ is obtained from a quite cumbersome six-parameters fitting of the experimental $V(z)$ to the relevant $\Upsilon$($z$) function.
The differential approach seems to be a plausible solution for a range of specific samples (of chemical or biological interests), but it requires the slowest possible measurement mode (the "full scan" method) and its feasibility had not been put to scrutiny above 2~kOe, that is well below the full extent of the magnetic field available for experiments.

In conclusion of this brief survey we note that neither of the approaches presented above fulfills the conditions required for precision magnetometry in the whole $T$ and $H$ ranges available in the MPMS magnetometers.
Below, we provide a full account of the \textsl{in situ} compensation \textsl{scheme} allowing the elimination of the magnetic signal of bulky substrates and other unwanted components, frequently disturbing the integral magnetometry of nanoscale magnetic materials.
In practical terms we expand and quantify the previous concept \cite{WangMu:2016_PRB} to the point that (i) makes routine measurements practically immune to the major systems instabilities, whereas, at the same time, (ii) permitting determination of absolute magnitudes of $m_{\mathrm{X}}$ even for samples of not strictly matching shapes and compositions to that of the compensating strips, all achieved with a better precision than in the conventional approach.

\section{Implementation of the method}

\subsection{Preparation of the sample holder for the \textsl{in situ} compensation}

\begin{figure}[t]
	\includegraphics[width=0.9\columnwidth]{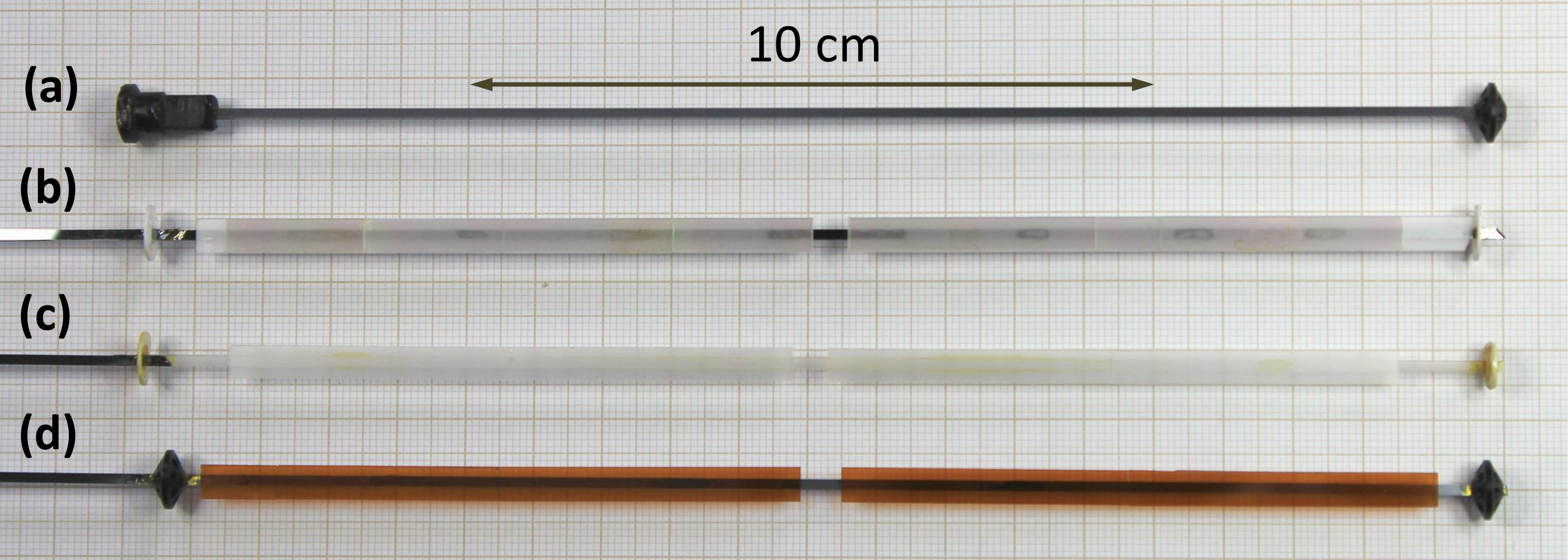}
	\caption{\label{fig:CSH}%
Examples of the sample holders made by the authors: (\textbf{a}) an example of a conventional one made of 1.5 mm wide and about 20 cm long Si strips, which also can be a base for (\textbf{b-d}) assemblies allowing \textsl{in situ} compensation of unwanted flux of either the substrate or a bulk part of the sample. They are customized to work with Quantum Design MPMS magnetometers.
In assemblies (b) and (d) the strips of the compensating materials (sapphire and GaP, respectively) differ from the material used for the supporting stick - Si, whereas in the assembly (c) all parts are made of sapphire. Most of the results presented in this report have been obtained using this particular sample holder. }%
\end{figure}

The actual form of the compensating sample holders (CSH) considered in this report are exemplified in Figure~\ref{fig:CSH}~(b-d).
Since our main experimental concern has been focused on thin (Ga,Mn)N layers, which are routinely, and most economically, deposited on sapphire substrates, the main designs contain the substrate-compensating strips cut from 2" sapphire wafers, panels (b) and (c) of Figure~\ref{fig:CSH}.
By the same token, for demanding studies of the magnetic layers grown on GaP substrates, another CSH presented in Figure~\ref{fig:CSH}~(d) has compensating strips cut from a 2" GaP wafer.
It is most advantageous to prepare the compensation strips from the same batch of the wafers on which the investigated material is grown.
This guarantees the closest possible levels of magnetic contamination present both in the investigated sample and in the compensating strips, and so a truly effective compensation.
Actually, when 2" wafers are available, four ~4.5 cm long strips are needed to be cut to provide nearly full coverage of the available space sideways of the sample (two on each side of the sample).
The mechanical constraints in MPMS SQUID magnetometers restrict the total length of the sample holder to less than 20 cm, and such a length, as established by numerical simulations similar to these presented in Fig.~\ref{fig:ES(ends)}, is more than sufficient to compensate most of the typically used diamagnetic substrates without introducing any experimentally noticeable modifications to the sample's $V(z)$, and so to established magnitudes of $m$.

The compensating strips are then glued in pairs to the supporting stick in such a way to form an opening (a gap) in the middle of the holder of a width, preferably, a bit greater than the typical width of the investigated samples.
This is the place where the investigated sample is placed.
Somewhat enlarged width of the gap actually easies both mounting and removing the sample from the opening, without compromising the overall performance of the CSH.
In fact, Eq.~(\ref{Eq:CQ}) given in section~\ref{sec:wzor} takes care of a lack of the perfect compensation, so a certain flexibility in this point will not do any harm.
In the first two examples given in Figure~\ref{fig:CSH} the gaps are about $\sim 5.2 \times 5$~mm$^2$, i.e. a fraction of a mm wider than the typical width of the (Ga,Mn)N samples investigated in the authors' lab.
The gap in the last CSH is wider due to different demands of GaP-based layers.
It turns out that the role of the material serving as the mechanical support for the whole assembly is of a prime importance if precise $T$-dependent studies are planned.
By far the most consistent performance is obtained when the supporting stick is made of the same material as the rest of the assembly.
The example of such a CSH is given in Figure~\ref{fig:CSH}~(c), where instead of Si, the most easily available, clean, and affordable material for this purpose, the supporting stick is made of a 2~mm wide and 200~mm long sapphire strip, cut from a $200 \times 100 \times 0.5$~mm$^3$ $R$-plane sapphire plate.
A strongly diluted GE varnish is used to firmly join the parts. 
The whole assemblies are finished with customized standard MPMS drinking straw adaptors [seen on the conventional sample holder in Figure~\ref{fig:CSH}~(a)], enabling an easy connection to the standard MPMS graphite probe.

The design suggested by the authors, although quite natural, turns out to be also quite robust.
Such assemblies as in Figure~\ref{fig:CSH} endure for quite a long time.
The silent feature of Figure~\ref{fig:Czas} is that the interrogated CSH assembly has survived intact more than two years while serving for more than a hundred measurement runs performed usually in the full envelope of the environmental parameters attainable in the MPMS magnetometer.

\subsection{Exemplary studies using CSH}
\label{sec:exemplary}

\begin{figure}[t]
	\includegraphics[width=0.95\columnwidth]{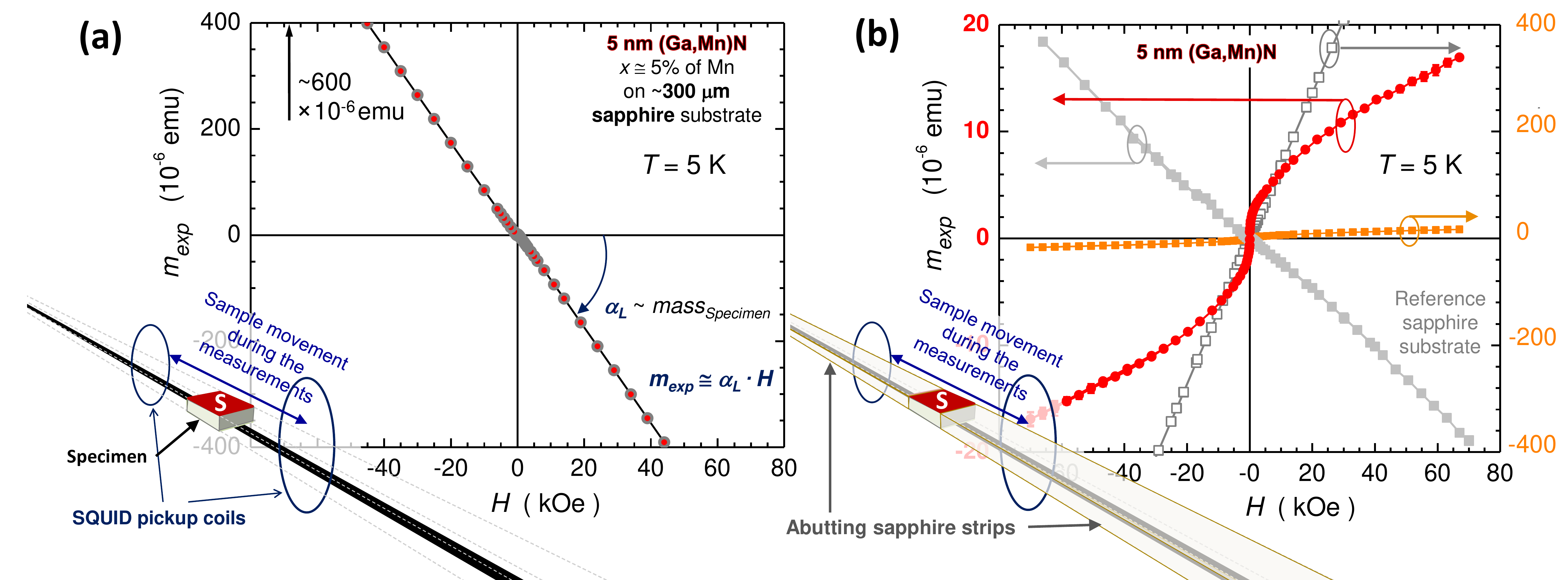}
	\caption{\label{fig:KX}%
A general outline of the \textsl{in situ} compensation method, applied to magnetic field $H$ dependent studies. The  specimen S investigated at $T = 5$~K  consists of a 5~nm thin (Ga,Mn)N layer (marked in red) grown on about 300~$\mu$m thick sapphire substrate (grey).
Both panels present the unprocessed magnetometer's output $m_{\mathrm{exp}}(H)$.
Panel \textbf{(a)} depicts $m_{\mathrm{exp}}(H)$  when S is mounted on a conventional sample holder (a long Si stick, black).
In this case the overwhelming majority of the flux contributing to $m_{\mathrm{exp}}(H)$ comes from the bulky (diamagnetic) substrate and so this part of $m_{\mathrm{exp}}(H)$ transmits all magnetometer performance flaws.
When the same specimen is mounted into a compensational sample holder (CSH), i.e. is abutted 
by strips of matching substrate material, panel \textbf{(b)}, then 
the substrate's flux is nearly nulled out, so the remaining net flux 
comes predominantly from the (magnetic) layer of interest (red bullets).
Note the 20 times expanded left Y scale with respect to (a).
For comparison orange bullets represent the same data set but plotted on the same scale as in panel (a) (right Y scale).
Light gray squares depict a reference sapphire sample measured in the same CSH, whereas the empty gray squares mark  $m_{\mathrm{exp}}(H)$ of the empty CSH - here the source of the flux is the gap between compensating strips.
}%
\end{figure}

The main experimental benefit of working with CSH is presented in Figure~\ref{fig:KX}, where we compare unprocessed results of $m(H)$ of the same $d \simeq 5$~nm, $x \simeq 5$\% (Ga,Mn)N layer measured using a conventional sample holder [as seen in Figure~\ref{fig:CSH}~(a)] and with the \textsl{in situ} compensation, that is using a CSH similar to that depicted in  Figure~\ref{fig:CSH}~(b).
These experimental configurations are sketched in the respective panels of Figure~\ref{fig:KX}.
Clearly, the abutting sapphire strips have reduced the flux of the substrate by a factor of $\sim 30$, which corresponds to mechanical thinning of the sapphire substrate down to about 10~$\mu$m - a process feasible but quite cumbersome due to the hardness of sapphire.
Importantly, this also means that the instabilities reported in Figs.~\ref{fig:1-SzBrA} and \ref{fig:Czas} have been reduced in the same proportion, \textsl{i.e.} below 0.02~$\mu$emu in this case, opening wide the door for the most demanding experimental research in this field.
Further examples of the qualitatively improved consistency of the results are given in Supplementary Figure S2.
Another important fact shown there is a five-fold reduction of the experimental $\sigma$, indicating that the \textsl{in situ} compensation reduces significantly the noise level associated with the reduction of $V(z)$ to $m$.

The data presented in Figure~\ref{fig:KX}~(b) indicate also that at such high levels of compensation ($\geq 97$\%) a nonlinear $m(H)$ of the (Ga,Mn)N layer is already clearly seen in the unprocessed results yielded by the magnetometer (red circles).
However, despite such a strong compensation this $m_{\mathrm{exp}}(H)$ does not show a clear tendency to saturation, yet.
Neither do the results obtained for a reference sapphire sample - marked by light-grey squares in the same panel.
Importantly, neither does the latter $m(H)$ null out.
This is an example of the typical experimental situation, when both the sample and its reference are of different dimensions and weights than is the opening (the gap) formed between the compensating strips.
In the case presented here the reference sapphire sample is about 3\% heavier whereas the sapphire substrate of the magnetic layer is about 1.6\% lighter than the gap.
So, it has to be said that the  $m_{\mathrm{S}}(H)$ and $m_{\mathrm{R}}(H)$  obtained in a CSH, apart from possessing visibly increased experimental fidelity, do not permit one to accurately  establish the magnitude of $m_{\mathrm{X}}(H)$.
Further data manipulation is required and  Eq.~(\ref{Eq:CQ})  introduced in the next section indeed allows to unambiguously establish the absolute magnitude of the $m_{\mathrm{X}}(H)$ specific to the investigated layer.
The final results are presented in Figure~\ref{fig:Examples}~(a).

The data selected for presentation in Figure~\ref{fig:KX} clearly demonstrate that even the preparation of a well-balanced CSH, in which the signals related to the magnetometer's instabilities would be reduced as much as 20 to 100 times, still does not guarantee an equivalent advancement in the absolute precision of the studies.
Following the examples given in Figure~\ref{fig:KX}~(b) we note that only in the idealistic case of truly 100\% compensation could a single measurement of a sample in a matching CSH be sufficient to establish the moment of the researched magnetic layer $m_{\mathrm{X}}$.
In this case $m_{\mathrm{X}} \equiv m_{\mathrm{exp}}$.
However, in the real case the long compensating strips are not cut from exactly the same source from which came the substrate of the investigated layer, and the whole CSH assembly is not free from minor structural imperfections.
Both these effects can exert a strong negative effect on the accuracy.

\subsection{Determination of the magnetic moment}
\label{sec:wzor}

In order to eliminate these spurious fluxes we recall the subtraction approach.
Assuming that both the sample and the reference are measured in \textsl{the same} CSH, then the difference of these moments
will be devoid of any common signals, particularly those specific to possible imperfections of CSH, and should yield directly the sought $m_{\mathrm{X}}$.
Since this is true for all types of measurements, to  make our considerations more general, we spell out this difference as: $L_{\mathrm{X}}$ = $L_{\mathrm{S}} - L_{\mathrm{R}}$,
where $L_{\mathrm{S}}$ and $L_{\mathrm{R}}$ stand for a full sets of relevant $T$-, $H$-, or time $t$- dependent measurements performed for the sample and the reference, respectively.
$L_{\mathrm{X}}$ denotes the set of results for the subject of the research.
Contrary to uncompensated measurements, the sought $L_X$ is established now as a difference of two rather small signals, i.e. comparable to the magnitude of the final $L_{\mathrm{X}}$, so the precision of the process is enhanced greatly.
Moreover, since both $L_{\mathrm{S}}$ and $L_{\mathrm{R}}$ are practically devoid of magnetometer instability signals -- their detrimental effect on $L_{\mathrm{X}}$ is also reduced considerably.
Furthermore, in the case considered here the difference $L_{\mathrm{S}} - L_{\mathrm{X}}$ is established accurately irrespective of the achieved compensation level.

However, it is practically impossible to have so well matched the sample and the reference, without destroying the former to obtain the latter.
Moreover, in everyday practice one reference specimen serves to provide the reference signal for a range of corresponding samples.
In this case the difference $L_{\mathrm{S}} - L_{\mathrm{R}}$ will contain also a small portion of the signal of the empty sample holder due to the different degrees of compensation. 
Therefore, a third measurement of the empty sample holder $L_{\mathrm{G}}$ is needed.
Practically, $L_{\mathrm{G}}$ is dominated by the signal exerted by the gap formed between the compensating strips (as presented in Fig.~\ref{fig:ES(ends)} and exemplified for the case considered in the previous section in Figure~\ref{fig:KX}~(b) by empty gray squares).
In order to quantify the magnitude of sample holder contribution in the case of the non-perfect compensation, $L_{\mathrm{G}}$ has to be scaled down by a factor corresponding to the degree of the "magnetic" filling of the gap by the inserted specimens: $\beta_i  =\mu_i \gamma_i/\mu_{\mathrm{G}} \gamma_{\mathrm{G}}$,
where 
$\mu_i$ and $\gamma_{\mathrm{i}}$ are the corresponding masses and size correcting factors for the two specimens and the gap.
The correcting factors $\gamma_i$ 
are needed to recover the proper magnitudes of $L_i$ 
in accordance with the real geometrical extent of the sources of the relevant fluxes.
This is because the MPMS output is provided in the point dipole approximation.
Following the notation adopted previously \cite{Sawicki:2011_SST}, MPMS output data have to be divided by the relevant $\gamma$ in order to recover the correct magnitude of $m$.
More about the role of $\gamma$ and its relevance is discussed in section S4 of the Supplementary Information.
Finally, we remark that $\mu_{\mathrm{G}}$, the mass of the missing material between the compensating strips, assumes \textsl{negative} magnitudes.
A method of its precise determination is given in Supplementary Figure S3.

Accordingly, the whole correction term taking care of the sample holder contribution to $L_{\mathrm{X}}$ for the case of non-identical shapes and masses of the sample and the reference assumes the form:
\begin{equation}\label{Eq:BQ}
\beta = \frac{\mu_{\mathrm{R}}\gamma_{\mathrm{R}} - \mu_{\mathrm{S}}\gamma_{\mathrm{S}}}{\mu_{\mathrm{G}} \gamma_{\mathrm{G}}},
\end{equation}
and the full expression for $L_{\mathrm{X}}$ is:
\begin{equation}\label{Eq:CQ}
L_{\mathrm{X}}\gamma_{\mathrm{X}} = L_{\mathrm{S}} - L_{\mathrm{R}} + \beta L_{\mathrm{G}}.
\end{equation}
The form in which Eq.~(\ref{Eq:CQ}) is given underlines an important fact that it is valid for all types of $T$-, $H$-, $t$- dependent studies performed within the whole envelope of $T$ and $H$ ranges available in the magnetometer.
Moreover, it also holds valid when the compensating strips and the substrates of the investigated samples are made of different materials.
The differences in the magnetic susceptibilities are accommodated in the experimentally established magnitudes of $L_i$.
So, the general form of the equation given here and the whole frame of the elaborated experimental method constitute a major qualitative advancement with respect to the other attempts \cite{Fukumura:2001_APL,Cabassi:2010_MST,WangMu:2016_PRB}.

Furthermore, as laboriously verified, all the necessary values of $L_i$ needed to obtain the correct magnitude of $L_{\mathrm{X}}$  come directly from the values reported by the MultiVu software in the corresponding *.dat files. 
It is worth stressing that there is no need to numerically simulate the expected shape of $\Upsilon$($z$) for each of these complicated experimental configurations.
Neither are any numerical fittings required to reduce the experimentally measured $V(z)$ into $m$.
However, the latter remark remains valid only as long as the CSH is properly centered.

The centering of the sample holder is one of the most important experiment-execution prerequisites, which have to be observed to achieve a truly sizable boost in accuracy and repeatability of the measurements.
Here we strongly advise the reader to perform the pre-measurement centering of the \textsl{empty} CSH always at the same $T$ and $H$.
The authors' choice has been 300~K and 20~kOe, respectively, the latter always ramped up from $H \simeq 0$.
Actually, the data presented in Figure~\ref{fig:Czas} have been collected during the pre-measurement centering of the empty CSH, and the added value brought about by such a consecutive recording of the signal of the bare CSH at the repeatable conditions is a long term monitoring of both of the sample holder and of the condition of the magnetometer.
Centering  issues, as well as some other practical details concerning the setting of the measurements are given in the Supplementary Information section "Setting of the measurements".
It is strongly suggested to collect all $L_i$ measurements along equivalent experimental sequences to assure the same thermal and magnetic history of the specimens and repeatable magnitudes of $H_{\mathrm{real}}$.

Finally, we remark that the workload required for this experimental approach is less than is dictated by Eq.~\ref{Eq:CQ}.
The weight of the $L_{\mathrm{G}}$ in the resulting magnitude of $L_{\mathrm{X}}$ is marginal and becomes the smaller the more the geometrical dimensions of the samples and the reference correspond to each other.
This causes $\beta \rightarrow 0$.
Therefore, it is sufficient to measure the full suite of $L_{\mathrm{G}}$ measurements once and repeat them rather infrequently, say even once a year, providing that neither the  experimental configuration nor the CSH itself  has  changed in the meantime.
For example, re-doing of the whole suite of $L_{\mathrm{G}}$ measurements, takes in particular care of the very weak trend down in time domain of the magnitude of the CHS response as can be noticed in Figure~\ref{fig:Czas}.
The same applies to $L_{\mathrm{R}}$, though in this case more frequent checks are advisable due to the equal importance of $L_{\mathrm{R}}$ and $L_{\mathrm{S}}$ in the determination of $L_{\mathrm{X}}$.
In any case, however, any newly performed  $T$-, $H$-, or $t$-dependence of the sample calls for well-matched measurements of $L_{\mathrm{ref}}$ and $L_{\mathrm{G}}$.
Therefore, this method works particularly well for a range of qualitatively similar samples for which the same suite of  $L_{\mathrm{R}}$ measurements is needed.
Then the main experimental effort can be directed to measure $L_{\mathrm{S}}$ only, as in the traditional approach to magnetometry.

\section{Examples of application}

As presented in Figure~\ref{fig:KX}~(b) a single measurement in a CSH is not sufficient to return the real magnitude of $m_{\mathrm{X}}$.
In that particular example the second measurement $m_{\mathrm{R}}$ of about 4.6\% heavier reference yields the opposite slope of its $m(H)$.
Therefore, their bare difference $L_{\mathrm{S}} - L_{\mathrm{R}}$ will yield $L_{\mathrm{X}}$ of an even greater slope, even if $L_{\mathrm{R}}$ is rescaled according to the relevant magnitudes of $\mu_{\mathrm{S}}\gamma_{\mathrm{S}}$ and $\mu_{\mathrm{R}}\gamma_{\mathrm{R}}$.
It is actually the $ \beta L_{\mathrm{G}}$ term in  Eq.~(\ref{Eq:CQ}), which, by bringing into consideration the extra flux generated by a not ideally filled gap of the CSH during   $L_{\mathrm{S}}$  and  $L_{\mathrm{R}}$ measurements,  restores the real magnitude of  $L_{\mathrm{X}}$, and eventually  $m_{\mathrm{X}}$.

\begin{figure}
	\includegraphics[width=0.9\columnwidth]{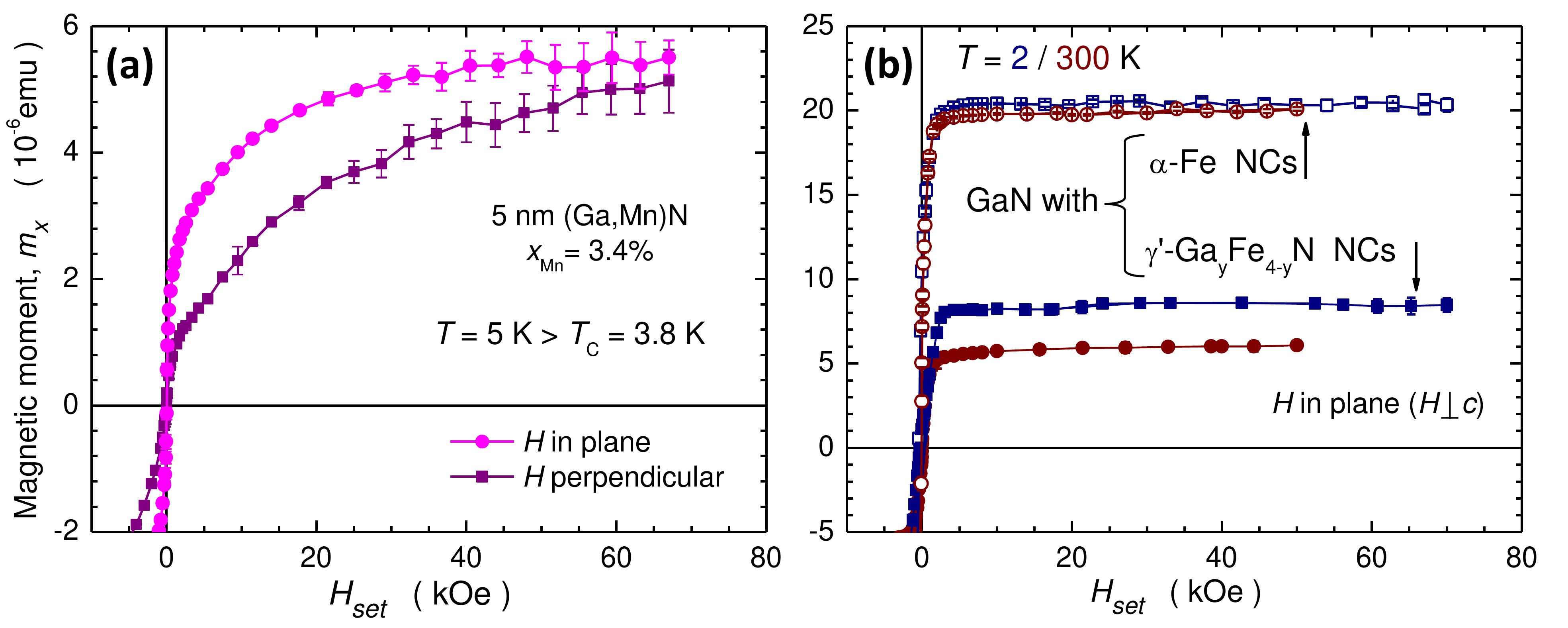}
	\caption{\label{fig:Examples}%
\textbf{(a)} Results of in- and out- of plane (bullets and squares, respectively) magnetic measurement of a 5~nm thin (Ga,Mn)N layer obtained using \textsl{in situ} compensation and the data reduction according to  Eq.~(\ref{Eq:CQ}).
In case of the in plane configuration the $m(H)$ of the sample, reference and the empty compensational sample holder (CSH) are presented in Figure~\ref{fig:KX}~(b).
\textbf{(b)} Magnetic responses corresponding to $\alpha$-Fe (open symbols) and $\gamma$'-Ga$_y$Fe$_{4-y}$N nanocrystals (closed symbols) embedded in a GaN matrix established at 2 and 300~K upon measurements in the same CSH as in (a) and data reduction according to Eq.~(\ref{Eq:CQ}).
}%
\end{figure}

The first example of the whole method in action is presented in Figure~\ref{fig:Examples}~(a), where magenta circles mark the results obtained by the application of Eq.~(\ref{Eq:CQ}) to the presented in Figure~\ref{fig:KX}~(b) set of $L_{\mathrm{S}}$, $L_{\mathrm{R}}$ and $L_{\mathrm{G}}$ measurements collected during the investigation of a nominally 5~nm thin (Ga,Mn)N layer of an expected Mn content between 3 and 5\%.
The correctness of the result is corroborated by the following two notions.
Firstly, even at the strongest available fields the established $m_{\mathrm{X}}(H)$ retains just a slightly convex character -- very alike to that of the previously investigated much thicker (Ga,Mn)N layers -- corresponding very closely at high field to the theoretical $m(H)$ for the Mn$^{3+}$ configuration in wurtzite GaN environment \cite{Stefanowicz:2010_PRB,Bonanni:2011_PRB,Sztenkiel:2016_NC}, yielding  the Mn content  $x$ = 3.4\%, a value close to the technological one.
Secondly, the another set of results in Figure~\ref{fig:Examples}~(a) (marked by purple squares) saturates at a very similar level.
This independently  performed measurement has been obtained for the same sample mounted in the perpendicular orientation (different $L_{\mathrm{S}}$ and  $\gamma_{\mathrm{S}}$, but the same  $\mu_{\mathrm{S}}$, $L_{\mathrm{R}}$ and $L_{\mathrm{G}}$ and their  $\mu_{\mathrm{i}}\gamma_{\mathrm{i}}$).
So, both measurements constitute together the first experimental assessment of the magnetic anisotropy in such a very thin ferromagnetic (Ga,Mn)N layers,
and this accomplishment would have not been possible without the elaboration of the \textsl{in situ} compensation method and the pertinent data reduction performed within the frame set by Eq.~(\ref{Eq:CQ}).

Another example,  where the precise knowledge of the absolute values of magnetic moment proves fundamental is during investigations of structures expected to contain an antiferromagnetic AFM component, which usually exhibits a marginally weak magnetic response.
In this view, CSHs have been used to study ensembles of Fe-rich, $\gamma$'-Ga$_y$Fe$_{4-y}$N nanometer-size nanocrystals (NCs) embedded in a GaN matrix, similar to those structurally characterized previously \cite{Navarro:2012_APL}.
An example of the magnetic properties of such ensembles is given in Figure~\ref{fig:Examples}~(b), where magnetic responses corresponding to $\alpha$-Fe NCs (open symbols) and $\gamma$'-Ga$_y$Fe$_{4-y}$N ones (closed symbols) measured at 2 and 300 K are plotted.
In this case it is not the establishment of the magnitude of the magnetic saturation of these two ensembles that defines the real technical merit of the result.
The magnetic saturation in these easy saturating magnetic systems can be equally accurately established by the traditional approach from the measurements performed at the weak field region ($H < 10$~kOe), i.e. where the flux of substrate is small comparably to that of the NCs and so the magnetometer instabilities do not mar the $m_{\mathrm{X}}$ of the NCs.
The real added value of the \textsl{in situ} compensation and the following data manipulation sets by Eq.~(\ref{Eq:CQ}) is the establishment that the magnetic response from these two NCs system does not show any high-field kink or an inclined $m(H)$ suggestive of a spin-flop transition or a spin canting, which are characteristic of AFM systems.
On the other hand, the saturation levels established in these measurements allow one to verify other magnetic characteristics of these prospective phase separated materials \cite{NavarroQ:2019_PRB}.

It has to be added, that the concept of the \textsl{in situ} compensation proves extremely useful in the studies of a broad range of systems.
As an example, we take the determination of the level of magnetically responsive contaminants in bulk materials.
In case referred to here the very same sample holder which is depicted in Figure~\ref{fig:CSH}~(b) was used to compensate the bulk diamagnetism in a search for possible superconducting precipitates in samples of topological Pb$_{0.20}$Sn$_{0.80}$Te and topologically trivial Pb$_{0.80}$Sn$_{0.20}$Te \cite{Mazur:2017_arXiv}.
By adjusting the masses of the studied Pb$_{1-y}$Sn$_y$Te samples, and after correcting for the low-$T$ sapphire response, the increased resolution of the method allowed to establish that the relative weight of precipitates that could produce a response specific to superconducting Pb or Sn was below 0.1~ppm, a level that goes beyond the state of the art of the integral magnetometry,
proving that this method can be employed widely as a very sensitive characterization tool in material science.

\section{Conclusions}

In this report a thorough method for mitigating signal instability problems in commercially available SQUID-based integral magnetometers has been put forward.
The method is based on the \textsl{in situ} magnetic compensation, at the sample holder level, of the vast majority of the dominating unwanted signal of the sample substrate, bulkiness or a carrier, which normally accompanies the minute object of interest.
Because the signal which is processed by the magnetometer  is typically up to two orders of magnitude smaller, the output is much less dependable on the inevitable fluctuations of some environmental variables, that otherwise detrimentally reduce the real credibility of the outcome in the standard approach to precision magnetometry.
In practice a two- to five-fold reduction in the absolute noise level has been observed.
Practical solutions for the achievement of adequate compensation and proper expressions to evaluate the final results obtained in the compensational sample holder are given.
The universal form of this expression allows to practically employ one design of the compensational sample holder in investigations of a range of specimens characterized by different sizes, shapes and compositions.
Importantly, the method does not require any involving modelling of the magnetometer output signal and laborious fitting.
All the required inputs to calculate the absolute magnitude of the net moment of the investigated object can be taken directly from the standard magnetometer output files.
The method has been implemented in MPMS SQUID magnetometers, but it is by no means limited to this particular system. It is exemplified and put to the test on nanometer thin layers of dilute magnetic semiconductors, without and with embedded nanocrystals.
The solution given here is of  great relevance to numerous fields of material science (to a broad community), where magnetic investigations are becoming  of  prime importance: biophysics, organic spintronics, and in further emerging new fields dealing with topological insulators, 3D Dirac semimetals, and 2D materials.

\section{Acknowledgements}

The authors acknowledge funding from the National Science Centre, Poland through FUGA Grant DEC-2014/12/S/ST3/00549 and OPUS Grants
DEC-2013/09/B/ST3/04175 and DEC-2017/27/B/ST3/02470, material issues assistance of Gerd Kunert, Alberta Bonanni, Vim van Roy, and Tomasz Baraniecki, and Andrea Navarro-Quezada (ANQ) for providing samples with embedded magnetic nanocrystals. The authors are further indebted to ANQ, and Andrew Rushford for critical reading of the manuscript.

\section{References}

\bibliography{ref_16Mar19}

\providecommand{\newblock}{}
\begin{thebibliography}{10}
\expandafter\ifx\csname url\endcsname\relax
  \def\url#1{{\tt #1}}\fi
\expandafter\ifx\csname urlprefix\endcsname\relax\def\urlprefix{URL }\fi
\providecommand{\eprint}[2][]{\url{#2}}

\bibitem{Ney:2001_EPL}
Ney A, Poulopoulos P and Baberschke K 2001 {\em Europhys. Lett.\/} {\bf 54} 820

\bibitem{Sawicki:2010_NP}
Sawicki M, Chiba D, Korbecka A, Nishitani Y, Majewski J~A, Matsukura F, Dietl T
  and Ohno H 2010 {\em Nat. Phys.\/} {\bf 6} 22

\bibitem{Chibal:2016_SR}
Chiba D, Shibata N and Tsukazaki A 2016 {\em Scientific Reports\/} {\bf 6}
  38005 \urlprefix\url{http://dx.doi.org/10.1038/srep38005}

\bibitem{Gladczuk:2017_JPDAP}
Gladczuk L, Lasek K, Puzniak R, Sawicki M, Aleshkevych P, Paszkowicz W,
  Minikayev R, Demchenko I~N, Syryanyy Y and Przyslupski P 2017 {\em Journal of
  Physics D: Applied Physics\/} {\bf 50} 485002
  \urlprefix\url{http://stacks.iop.org/0022-3727/50/i=48/a=485002}

\bibitem{Hayassi:2018_APE}
Hayashi Y, Hibino Y, Matsukura F, Miwa K, Ono S, Hirai T, Koyama T, Ohno H and
  Chiba D 2018 {\em Applied Physics Express\/} {\bf 11} 013003
  \urlprefix\url{http://stacks.iop.org/1882-0786/11/i=1/a=013003}

\bibitem{Sawicki:2018_PRB}
Sawicki M, Proselkov O, Sliwa C, Aleshkevych P, Domagala J~Z, Sadowski J and
  Dietl T 2018 {\em Phys. Rev. B\/} {\bf 97}(18) 184403
  \urlprefix\url{https://link.aps.org/doi/10.1103/PhysRevB.97.184403}

\bibitem{Sandaresan:2006_PRB}
Sundaresan A, Bhargavi R, Rangarajan N, Siddesh U and Rao C~N~R 2006 {\em Phys.
  Rev. B\/} {\bf 74}(16) 161306
  \urlprefix\url{https://link.aps.org/doi/10.1103/PhysRevB.74.161306}

\bibitem{Sadowski:2011_PRB}
Sadowski J, Domagala J~Z, Mathieu R, Kov\'acs A, Kasama T, Dunin-Borkowski R~E
  and Dietl T 2011 {\em Phys. Rev. B\/} {\bf 84} 245306

\bibitem{Sueli:2016_JPCC}
Masunaga S~H, Jardim R~F, Correia M~J and Figueiredo W 2016 {\em The Journal of
  Physical Chemistry C\/} {\bf 120} 765--770 (\textit{Preprint}
  \eprint{https://doi.org/10.1021/acs.jpcc.5b10933})
  \urlprefix\url{https://doi.org/10.1021/acs.jpcc.5b10933}

\bibitem{Rath:2011_JMMM}
Rath C, Mohanty P and Banerjee A 2011 {\em Journal of Magnetism and Magnetic
  Materials\/} {\bf 323} 1698 -- 1702 ISSN 0304-8853
  \urlprefix\url{http://www.sciencedirect.com/science/article/pii/S0304885311000588}

\bibitem{Augustyns:2017_PRB}
Augustyns V, van Stiphout K, Joly V, Lima T~A~L, Lippertz G, Trekels M,
  Men\'endez E, Kremer F, Wahl U, Costa A~R~G, Correia J~G, Banerjee D,
  Gunnlaugsson H~P, von Bardeleben J, Vickridge I, Van~Bael M~J, Hadermann J,
  Ara\'ujo J~P, Temst K, Vantomme A and Pereira L~M~C 2017 {\em Phys. Rev. B\/}
  {\bf 96}(17) 174410
  \urlprefix\url{https://link.aps.org/doi/10.1103/PhysRevB.96.174410}

\bibitem{Siusys:2014_NL}
Siusys A, Sadowski J, Sawicki M, Kret S, Wojciechowski T, Gas K, Szuszkiewicz
  W, Kaminska A and Story T 2014 {\em Nano Letters\/} {\bf 14} 4263--4272

\bibitem{Sadowski:2017_Nanoscale}
Sadowski J, Kret S, Siusys A, Wojciechowski T, Gas K, Islam M~F, Canali C~M and
  Sawicki M 2017 {\em Nanoscale\/} {\bf 9}(6) 2129--2137
  \urlprefix\url{http://dx.doi.org/10.1039/C6NR08070G}

\bibitem{Peddis:2008_JPCC}
Peddis D, Cannas C, Musinu A and Piccaluga G 2008 {\em The Journal of Physical
  Chemistry C\/} {\bf 112} 5141--5147
  \urlprefix\url{https://doi.org/10.1021/jp076704d}

\bibitem{Sun:2017_QM}
Sun Y, Zheng Y, Pan H, Chen J, Zhang W, Fu L, Zhang K, Tang N and Du Y 2017
  {\em Quantum Materials\/} {\bf 2} 5
  \urlprefix\url{https://doi.org/10.1038/s41535-017-0010-2}

\bibitem{LiuYuan:2013_SR}
Liu Y, Tang N, Wan X, Feng Q, Li M, Xu Q, Liu F and Du Y 2013 {\em Scientific
  Reports\/} {\bf 3} 2566 \urlprefix\url{http://dx.doi.org/10.1038/srep02566}

\bibitem{LiuYuan:2016_NC}
Liu Y, Shen Y, Sun L, Li J, Liu C, Ren W, Li F, Gao L, Chen J, Liu F, Sun Y,
  Tang N, Cheng H~M and Du Y 2016 {\em Nature Communications\/} {\bf 7} 10921
  \urlprefix\url{http://dx.doi.org/10.1038/ncomms10921}

\bibitem{Pereira:2011_JPCM}
Pereira L~M~C, Som T, Demeulemeester J, Bael M~J~V, Temst K and Vantomme A 2011
  {\em Journal of Physics: Condensed Matter\/} {\bf 23} 346004
  \urlprefix\url{http://stacks.iop.org/0953-8984/23/i=34/a=346004}

\bibitem{Sawicki:2013_PRB}
Sawicki M, Guziewicz E, \L{}ukasiewicz M~I, Proselkov O, Kowalik I~A, Lisowski
  W, Dluzewski P, Wittlin A, Jaworski M, Wolska A, Paszkowicz W, Jakiela R,
  Witkowski B~S, Wachnicki L, Klepka M~T, Luque F~J, Arvanitis D, Sobczak J~W,
  Krawczyk M, Jablonski A, Stefanowicz W, Sztenkiel D, Godlewski M and Dietl T
  2013 {\em Phys. Rev. B\/} {\bf 88}(8) 085204
  \urlprefix\url{https://link.aps.org/doi/10.1103/PhysRevB.88.085204}

\bibitem{Pereira:2013_JPCM}
Pereira L~M~C, Wahl U, Correia J~G, Bael M~J~V, Temst K, Vantomme A and Araújo
  J~P 2013 {\em Journal of Physics: Condensed Matter\/} {\bf 25} 416001
  \urlprefix\url{http://stacks.iop.org/0953-8984/25/i=41/a=416001}

\bibitem{Henne:2016_PRB}
Henne B, Ney V, de~Souza M and Ney A 2016 {\em Phys. Rev. B\/} {\bf 93}(14)
  144406 \urlprefix\url{https://link.aps.org/doi/10.1103/PhysRevB.93.144406}

\bibitem{NeyV:2016_PRB}
Ney V, Henne B, Lumetzberger J, Wilhelm F, Ollefs K, Rogalev A, Kovacs A,
  Kieschnick M and Ney A 2016 {\em Phys. Rev. B\/} {\bf 94}(22) 224405
  \urlprefix\url{https://link.aps.org/doi/10.1103/PhysRevB.94.224405}

\bibitem{Sawicki:2006_JMMM}
Sawicki M 2006 {\em J. Mag. Magn. Mater.\/} {\bf 300} 1

\bibitem{Gas:2018_JALLCOM}
Gas K, Domagala J~Z, Jakiela R, Kunert G, Dluzewski P, Piskorska-Hommel E,
  Paszkowicz W, Sztenkiel D, Winiarski M~J, Kowalska D, Szukiewicz R,
  Baraniecki T, Miszczuk A, Hommel D and Sawicki M 2018 {\em Journal of Alloys
  and Compounds\/} {\bf 747} 946 -- 959 ISSN 0925-8388
  \urlprefix\url{http://www.sciencedirect.com/science/article/pii/S0925838818309150}

\bibitem{Bujak:2013_CSR}
Bujak P, Kulszewicz-Bajer I, Zagorska M, Maurel V, Wielgus I and Pron A 2013
  {\em Chem. Soc. Rev.\/} {\bf 42}(23) 8895--8999
  \urlprefix\url{http://dx.doi.org/10.1039/C3CS60257E}

\bibitem{Kopani:2015_BioMetals}
Kop{\'a}ni M, Miglierini M, Lan{\v{c}}ok A, Dekan J, {\v{C}}aplovicov{\'a} M,
  Jakubovsk{\'y} J, Bo{\v{c}}a R and Mrazova H 2015 {\em BioMetals\/} {\bf 28}
  913--928 ISSN 1572-8773
  \urlprefix\url{https://doi.org/10.1007/s10534-015-9876-2}

\bibitem{Zhao:2014_NM}
Zhao L, Deng H, Korzhovska I, Chen Z, Konczykowski M, Hruban A, Oganesyan V and
  Krusin-Elbaum L 2014 {\em Nature Materials\/} {\bf 13} 580
  \urlprefix\url{http://dx.doi.org/10.1038/nmat3962}

\bibitem{Dutta:2017_SR}
Dutta P, Pariari A and Mandal P 2017 {\em Scientific Reports\/} {\bf 7} 4883
  \urlprefix\url{https://doi.org/10.1038/s41598-017-05164-9}

\bibitem{Abraham:2005_APL}
Abraham D~W, Frank M~M and Guha S 2005 {\em Appl. Phys. Lett.\/} {\bf 87}
  252502

\bibitem{Garcia:2009_JAP}
Garcia M~A, Pinel E~F, de~la Venta J, Quesada A, Bouzas V, Fern\'{a}ndez J~F,
  Romero J~J, Gonz\'{a}lez M~S~M and Costa-Kr\"{a}mer J~L 2009 {\em Journal of
  Applied Physics\/} {\bf 105} 013925 (pages~7)
  \urlprefix\url{http://link.aip.org/link/?JAP/105/013925/1}

\bibitem{Pereira:2011_JPDAP}
Pereira L~M~C, Araújo J~P, Bael M~J~V, Temst K and Vantomme A 2011 {\em
  Journal of Physics D: Applied Physics\/} {\bf 44} 215001
  \urlprefix\url{http://stacks.iop.org/0022-3727/44/i=21/a=215001}

\bibitem{Sawicki:2011_SST}
Sawicki M, Stefanowicz W and Ney A 2011 {\em Semicon. Sci. Technol.\/} {\bf 26}
  064006

\bibitem{Ney:2011_SST}
Ney A 2011 {\em Semiconductor Science and Technology\/} {\bf 26} 064010
  \urlprefix\url{http://stacks.iop.org/0268-1242/26/i=6/a=064010}

\bibitem{Ney:2008_JMMM}
Ney A, Kammermeier T, Ney V, Ollefs K and Ye S 2008 {\em J. Magn. Magn.
  Mater.\/} {\bf 320} 3341

\bibitem{Pereira:2017_JPDAP}
Pereira L~M~C 2017 {\em Journal of Physics D: Applied Physics\/} {\bf 50}
  393002 \urlprefix\url{http://stacks.iop.org/0022-3727/50/i=39/a=393002}

\bibitem{Buchner:2018_APL}
Buchner M, H{\"o}fler K, Henne B, Ney V and Ney A 2018 {\em Journal of Applied
  Physics\/} {\bf 124} 161101 (\textit{Preprint}
  \eprint{https://doi.org/10.1063/1.5045299})
  \urlprefix\url{https://doi.org/10.1063/1.5045299}

\bibitem{Stefanowicz:2013_PRB}
Stefanowicz S, Kunert G, Simserides C, Majewski J~A, Stefanowicz W, Kruse C,
  Figge S, Li T, Jakie{\l}a R, Trohidou K~N, Bonanni A, Hommel D, Sawicki M and
  Dietl T 2013 {\em Phys. Rev. B\/} {\bf 88} 081201(R)

\bibitem{WangMu:2016_PRB}
Wang M, Marshall R~A, Edmonds K~W, Rushforth A~W, Campion R~P and Gallagher B~L
  2016 {\em Phys. Rev. B\/} {\bf 93}(18) 184417
  \urlprefix\url{https://link.aps.org/doi/10.1103/PhysRevB.93.184417}

\bibitem{Stamenov:2006_RSI}
Stamenov P and Coey J~M~D 2006 {\em Review of Scientific Instruments\/} {\bf
  77} 015106 (\textit{Preprint} \eprint{https://doi.org/10.1063/1.2149190})
  \urlprefix\url{https://doi.org/10.1063/1.2149190}

\bibitem{Stefanowicz:2014_PRB}
Stefanowicz W, Adhikari R, Andrearczyk T, Faina B, Sawicki M, Majewski J~A,
  Dietl T and Bonanni A 2014 {\em Phys. Rev. B\/} {\bf 89}(20) 205201

\bibitem{Adhikari:2016_PRB}
Adhikari R, Matzer M, Mart\'{\i}n-Luengo A~T, Scharber M~C and Bonanni A 2016
  {\em Phys. Rev. B\/} {\bf 94}(8) 085205
  \urlprefix\url{https://link.aps.org/doi/10.1103/PhysRevB.94.085205}

\bibitem{Kunert:2012_APL}
Kunert G, Dobkowska S, Li T, Reuther H, Kruse C, Figge S, Jakiela R, Bonanni A,
  Grenzer J, Stefanowicz W, Borany J~v, Sawicki M, Dietl T and Hommel D 2012
  {\em Appl. Phys. Lett.\/} {\bf 101} 022413

\bibitem{Sawicki:2004_ICPS}
Sawicki M, Dietl T, Foxon C~T, Novikov S~V, Campion R~P, Edmonds K~W, Wang K~Y,
  Giddings A~D and Gallagher B~L 2005 Search for hole mediated ferromagnetism
  in cubic {(Ga,Mn)N} {\em Proceedings of the 27th International Conference on
  the Physics of Semiconductors\/} ({\em AIP Conference Proceedings\/} vol 772)
  ed de~Walle J~M~C~G~V p 1371

\bibitem{Sawicki:2012_PRB}
Sawicki M, Devillers T, Ga{\l}\c{e}ski S, Simserides C, Dobkowska S, Faina B,
  Grois A, Navarro-Quezada A, Trohidou K~N, Majewski J~A, Dietl T and Bonanni A
  2012 {\em Phys. Rev. B\/} {\bf 85} 205204
  \urlprefix\url{http://link.aps.org/doi/10.1103/PhysRevB.85.205204}

\bibitem{Stefanowicz:2010_PRB}
Stefanowicz W, Sztenkiel D, Faina B, Grois A, Rovezzi M, Devillers T,
  Navarro-Quezada A, Li T, Jakie{\l}a R, Sawicki M, Dietl T and Bonanni A 2010
  {\em Phys. Rev. B\/} {\bf 81} 235210

\bibitem{Bonanni:2011_PRB}
Bonanni A, Sawicki M, Devillers T, Stefanowicz W, Faina B, Li T, Winkler T~E,
  Sztenkiel D, Navarro-Quezada A, Rovezzi M, Jakie{\l}a R, Grois A, Wegscheider
  M, Jantsch W, Suffczy\'nski J, D'Acapito F, Meingast A, Kothleitner G and
  Dietl T 2011 {\em Phys. Rev. B\/} {\bf 84} 035206

\bibitem{Sztenkiel:2016_NC}
Sztenkiel D, Foltyn M, Mazur G~P, Adhikari R, Kosiel K, Gas K, Zgirski M,
  Kruszka R, Jakiela R, Li T, Piotrowska A, Bonnanni A, Sawicki M and Dietl T
  2016 {\em Nature Communications\/} {\bf 7} 13232
  \urlprefix\url{http://dx.doi.org/10.1038/ncomms13232}

\bibitem{QD1070-207-MST}
{Quantum Design} 2009 {\em MPMS Application Note 1070-207: Using PPMS
  Superconducting Magnets at Low Fields.\/}

\bibitem{QD1014-213-MST}
{Quantum Design} 2002 {\em MPMS Application Note 1014-213: Subtracting the
  Sample Holder Background from Dilute Samples.\/}

\bibitem{Fukumura:2001_APL}
Fukumura T, Jin Z, Kawasaki M, Shono T, Hasegawa T, Koshihara S and Koinuma H
  2001 {\em Appl. Phys. Lett.\/} {\bf 78} 958

\bibitem{Hayden:2017_RSI}
Hayden M~E, Lambinet V, Gomis S and Gries G 2017 {\em Review of Scientific
  Instruments\/} {\bf 88} 056106 (\textit{Preprint}
  \eprint{https://doi.org/10.1063/1.4983777})
  \urlprefix\url{https://doi.org/10.1063/1.4983777}

\bibitem{Zieba:1993_RSI}
Zieba A 1993 {\em Rev. Sci. Instrum.\/} {\bf 64} 3357

\bibitem{Miller:1996_RSI}
Miller L~L 1996 {\em Rev. Sci. Instrum.\/} {\bf 67} 3201

\bibitem{Cabassi:2010_MST}
Cabassi R, Bolzoni F and Casoli F 2010 {\em Measurement Science and
  Technology\/} {\bf 21} 035701
  \urlprefix\url{http://stacks.iop.org/0957-0233/21/i=3/a=035701}

\bibitem{Navarro:2012_APL}
Navarro-Quezada A, Devillers T, Li T and Bonanni A 2012 {\em Appl. Phys.
  Lett.\/} {\bf 101} 081911

\bibitem{NavarroQ:2019_PRB}
Navarro-Quezada A, Aiglinger M, Faina B, Gas K, Matzer M, Li T, Adhikari R,
  Sawicki M and Bonanni A 2019 {\em Phys. Rev. B\/} {\bf 99}(8) 085201
  \urlprefix\url{https://link.aps.org/doi/10.1103/PhysRevB.99.085201}

\bibitem{Mazur:2017_arXiv}
Mazur G, Dybko K, Szczerbakow A, Zgirski M, Lusakowska E, Kret S, Korczak J,
  Story T, Sawicki M and Dietl T 2017 {\em cond-mat\/}  arXiv:1709.04000

\end{thebibliography}

\end{document}